\documentclass[prl,twocolumn,showpacs,showkeys,superscriptaddress]{revtex4}
\usepackage{graphicx}
\def\be{\begin{eqnarray}}
\def\ee{\end{eqnarray}}
\def\om{\omega}

\def\prt{\partial}
\def\re{{\rm Re}}
\def\im{{\rm Im}}
\def\Re{\re}
\def\Im{\im}

\begin{document}
\title{Chiral Dynamics of Deeply Bound Pionic
Atoms}
\author{E.E.~Kolomeitsev}
\affiliation{ECT*, Villa Tambosi, I-38050 Villazzano
(Trento),Italy}
\author{N.~Kaiser}
\affiliation{Physik-Department, Technische Universit\"at
M\"unchen, D-85747 Garching, Germany}
\author{W.~Weise}
\affiliation{ECT*, Villa Tambosi, I-38050 Villazzano
(Trento),Italy} \affiliation{Physik-Department, Technische
Universit\"at M\"unchen, D-85747 Garching, Germany}

\begin{abstract}
We present and discuss a systematic calculation, based on
two--loop chiral perturbation theory, of the pion-nuclear s-wave
optical potential. A proper treatment of the explicit energy
dependence of the off-shell pion self-energy together with
(electromagnetic) gauge invariance of the Klein--Gordon equation
turns out to be crucial. Accurate data for the binding energies
and widths of the $1s$ and $2p$ levels in pionic  $^{205}$Pb and
$^{207}$Pb are well reproduced, and the notorious "missing
repulsion" in the pion-nuclear s-wave optical potential is
accounted for. The connection with the in-medium change of the
pion decay constant is clarified.
\end{abstract}
\maketitle

Recent accurate data on $1s$ and $2p$ states of
negatively charged pions bound to Pb nuclei \cite{pb207} have set
new standards and constraints for  the detailed analysis of s-wave
pion-nucleus interactions. This subject has a long history \cite
{ee66,Tauscher71,EW88,mstv90,nieves93,WBW97,BFG97}, culminating
in various attempts to understand the notorious "missing
repulsion": the standard ansatz for the s-wave pion-nucleus
optical potential in terms of the empirical threshold $\pi
N$-amplitudes times the proton and neutron densities $\rho_{p,n
}$, supplemented by sizeable double-scattering corrections,
still misses the empirically required repulsive
interaction by a large amount. On purely phenomenological grounds,
this problem can be fixed by simply introducing a sufficiently
large negative real part, $\Re B_0$, in the $\rho^2$ term of the
pion-nuclear optical potential~\cite{nieves93}. The arbitrariness
of this procedure is of course unsatisfactory, also in view of the
fact that $\Re B_0$ cannot be calculated microscopically with
sufficient accuracy, and that the high precision measurements of
the real part of the pion deuteron scattering length
\cite{pidscat} suggests a very small net $\Re B_0$ by analogy.

In the meantime, this issue has been revived in a  variety of
ways. An interpretation of the "missing repulsion" in terms of a
possible in-medium change of the pion decay constant, as suggested
in ref.~\cite{Weise01}, appears to be remarkably successful~\cite{KienYamaz02,Friedman02}
but has been debated~\cite{Wirzba02,Oset02}.
Clearly, this concept needs further
justification. In the present paper we show that a key to the
understanding of low-energy pion-nucleus s-wave interactions lies
in its distinct energy dependence imposed by chiral symmetry, in
combination with the "accidental" approximate vanishing of the
isospin-even threshold $\pi N$-amplitude.
Another important feature, generally ignored in previous
analyses, is the systematic incorporation of gauge invariance at
all places where the pion energy $\omega$ appears explicitly, when
solving the Klein-Gordon equation in the presence of
electromagnetic interactions.

Our framework will be in-medium chiral perturbation theory at
two-loop order~\cite{Kaiser01}. But before going into technical
details, the following  simplified treatment may be useful
to illustrate driving mechanisms.

Consider a zero-momentum  $\pi^-$ interacting with nuclear matter at
low proton and neutron densities, $\rho_p$ and $\rho_n$. The
in-medium pion polarization operator to leading
order in the nucleon densities is expressed in terms  of the
isospin-even and isospin-odd off-shell $\pi N$  amplitudes
$T^\pm(\omega)$ as $\Pi(\om)= -T^-(\omega)\, \delta \rho
-T^+(\omega)\, \rho$, with the isoscalar and isovector nucleon
densities, $\rho=\rho_p +\rho_n$ and  $\delta \rho
=\rho_p-\rho_n$.
The spontaneous and explicit breaking of chiral symmetry
implies the following leading terms of those amplitudes~\cite{Ericson87}:
\be\label{tpl}
T^+(\om)=\frac{\sigma_N-\beta\,\om^2}{f_\pi^2}\,,\quad
T^-(\om)=\frac{\om}{2f_\pi^2}\,\,, \ee
where $f_\pi=92.4$~MeV is the pion decay constant,
$m_\pi=139.57$~MeV is the (charged) pion mass and $\sigma_N\simeq
45\pm 8$~MeV~\cite{sainio}  is the pion-nucleon sigma-term.
The empirical observation that
$T^+(m_\pi)=(-0.04 \pm0.09)\,{\rm fm}\simeq  0$ \cite{Schroeder99}
sets the constraint $\beta\simeq \sigma_N/m_\pi^2$.

Next, consider the pionic in-medium dispersion equation at zero momentum,
$\om^2-m_\pi^2-\Pi(\om)=0$\,. Introduce an (energy independent) equivalent
optical potential $U$ by $\om^2=m_\pi^2+2\, m_\pi\, U$ with $U\ll
m_\pi$, and
expand around $\om=m_\pi$.
This gives
\be\label{uopt}
 U \simeq {\Pi(m_\pi) \over 2m_\pi} \bigg[ 1+ {\Pi'(m_\pi)
\over 2m_\pi} \bigg]+\dots
\simeq
\frac{\Pi(m_\pi)/(2m_\pi)}
{\left(1-\frac{\prt \Pi}{\prt \om^2}\right)\big|_{\om=m_\pi}}\,,
\ee
where the last approximate step introduces
the wave function renormalization factor
$\left(1-\frac{\prt \Pi}{\prt\om^2}\right)_{\om=m_\pi}^{-1}$\,.
The difference between second and last step in (\ref{uopt})
is of sub-leading order.
Inserting (\ref{tpl}) and assuming
$\delta\rho\ll \rho$  one finds
\be\label{ufpi}
U\simeq
-\frac{\delta \rho}{4\,f_\pi^2}\,
\left(1-\frac{\sigma_N \,
\rho}{m_\pi^2\,f_\pi^2}\right)^{-1}
\simeq-\frac{\delta \rho}{4\,f_\pi^{*2}}\,,
\ee
which is the expression proposed in
ref.~\cite{Weise01}.
It involves  the driving Weinberg-Tomozawa
term in $T^-(\om)$, but with the pion decay constant renormalized
($f_\pi\to f_\pi^*(\rho)$) to leading order in the density $\rho$,
in accordance with the corresponding in-medium change of the
chiral quark condensate $<\bar q\ q>$\,.

To the extent that  $U$ represents part of the (energy
independent) s-wave optical potential commonly used in the
phenomenological analysis of pionic atoms, at least part of the
"missing repulsion" is thus given a physical interpretation in
terms of the reduced in-medium $f_\pi^*$ in the denominator of
(\ref{ufpi}). Of course, rather than constructing the potential
$U$ and following the steps leading to (\ref{uopt},\ref{ufpi}),
one can directly solve the
Klein-Gordon (KG) equation with the full energy dependence of the
polarization operator $\Pi(\om)$. This is the procedure
systematically applied in this paper, with  proper recognition
of gauge invariance in the presence of the electromagnetic field.

The KG equation with Coulomb potential
$V_c(\vec r)<0$ and total pion self-energy, $\Pi_{\rm
tot}(\omega,\vec r\,)$  reads
\be\label{KGE}
\left[(\om-V_c)^2 +\nabla^2-m_\pi^2-\Pi_{\rm
tot}(\om-V_c,\vec r\,)\right]\,
\Phi(\vec{r}\,)=0\,. \ee
The total polarization operator expressed in terms of local proton
and neutron densities, $\Pi_{\rm tot}(\omega,\vec r\,)=
\Pi_{\rm tot}(\omega;\rho_p(\vec r),\rho_n(\vec r\, ))$, can be split into
its  s-wave and p-wave parts
\be\nonumber
\Pi_{\rm tot}(\omega;\rho_p,\rho_n)
=\Pi(\omega)
+ \Delta \Pi_{\rm S}(\omega; \rho_p, \rho_n)
+ \Pi_{\rm P}(\omega;\rho_p,\rho_n)\,,
\ee
where we
separate explicitly the phenomenological s-wave absorption term
quadratic in densities,
\be
\Delta \Pi_{\rm S}(\omega;\rho_p,\rho_n)=-8\pi
\left(1+\frac{m_\pi}{2M}\right) B_0\, \rho_p(\rho_n+\rho_p)\,,
\ee
parameterized as in ref.~\cite{nieves93}.
Here $M$ stands for the nucleon mass.
We use $\Im
B_0=0.063~m_\pi^{-4}$ and $\Re B_0=0$ as our standard set and
discuss variations of $\Re B_0$ later.
For the p-wave  part
$\Pi_{\rm P}(\omega;\rho_p,\rho_n)$ we use the traditional
Kisslinger form with inclusion of short-range correlations and
parameters as specified in  ref.~\cite{EW88} (set A).

Given the smallness of the isospin-even $\pi N$ scattering
amplitude $T^+(\omega)$, double scattering (Pauli-blocking)
corrections in $\Pi(\omega)$ are well known to be important
\cite{ee66}. When those are included the "phenomenological" s-wave
pion polarization operator becomes~\cite{WBW97}:
\be\label{piphen}
\Pi_{\rm phen}(\om;\rho_p,\rho_n)=-T^-(\om)\, \delta \rho -
T^+_{\rm eff}(\om,\rho)\, \rho\,, \ee
with
\be\label{teff}
T^+_{\rm eff}(\om,\rho)=T^+(\om)-\frac{3k_F}{8
\pi^2}\,\left([T^+(\om)]^2+2\,[T^-(\om)]^2\right)\,.
\ee
The local Fermi momentum $k_{F}(r)=[3\pi^2\rho(r)/2]^{1/3}$ is
rewritten in terms of the local density $\rho(r)$.
Taking the polarization operator  (\ref{piphen}) at
the on-shell pion energy $\omega = m_\pi$,
\be\label{pistat}
\Pi(\om) =\Pi_{\rm phen}(\om=m_\pi;\rho_p,\rho_n)\,,
\ee
we recover the traditional form
of the (energy independent) s-wave optical potential
\cite{EW88,BFG97}.
The proton and neutron density distributions
$\rho_p(r)$ and $\rho_n(r)$ are given as two-parameter Fermi
functions $\rho_j(r)=\rho_{0,j}[1+\exp((r-R_j)/a_{j})]^{-1}$. The
central density $\rho_{0,j}$ is normalized to the total number of
protons and neutrons in the nucleus. The proton radii, $R_p$, are
extracted from the nuclear charge radii following from the
analyses of muonic atoms \cite{Fricke95}, taking into account the
finite proton size $<r_p^2>=0.73$~fm$^2$: $R_p[^{205}{\rm
Pb}]=6.66$~fm and $R_p[^{207}{\rm Pb}]=6.67$~fm. Since the charge
radii have not been  measured for the complete chain of Pb
isotopes, we have  interpolated linearly between two neighboring
measured isotopes. The diffuseness coefficient is taken the same
for $^{205,207}$Pb,  $a_{p}=0.48$~fm. For the neutron radii we use
values from the proton-neutron rms-radius difference as obtained
in the Brueckner-Hartree-Fock calculations of ref.\cite{Angeli80}:
$R_{n}[^{205}{\rm Pb}]=6.94$~fm and $R_{n}[^{207}{\rm
Pb}]=6.97$~fm. We assume $a_n=a_p$. The numerical input is close
to that in refs.~\cite{Tauscher71,WBW97}.

\begin{figure*}
\begin{center}
\includegraphics[width=5.4cm,clip=true]{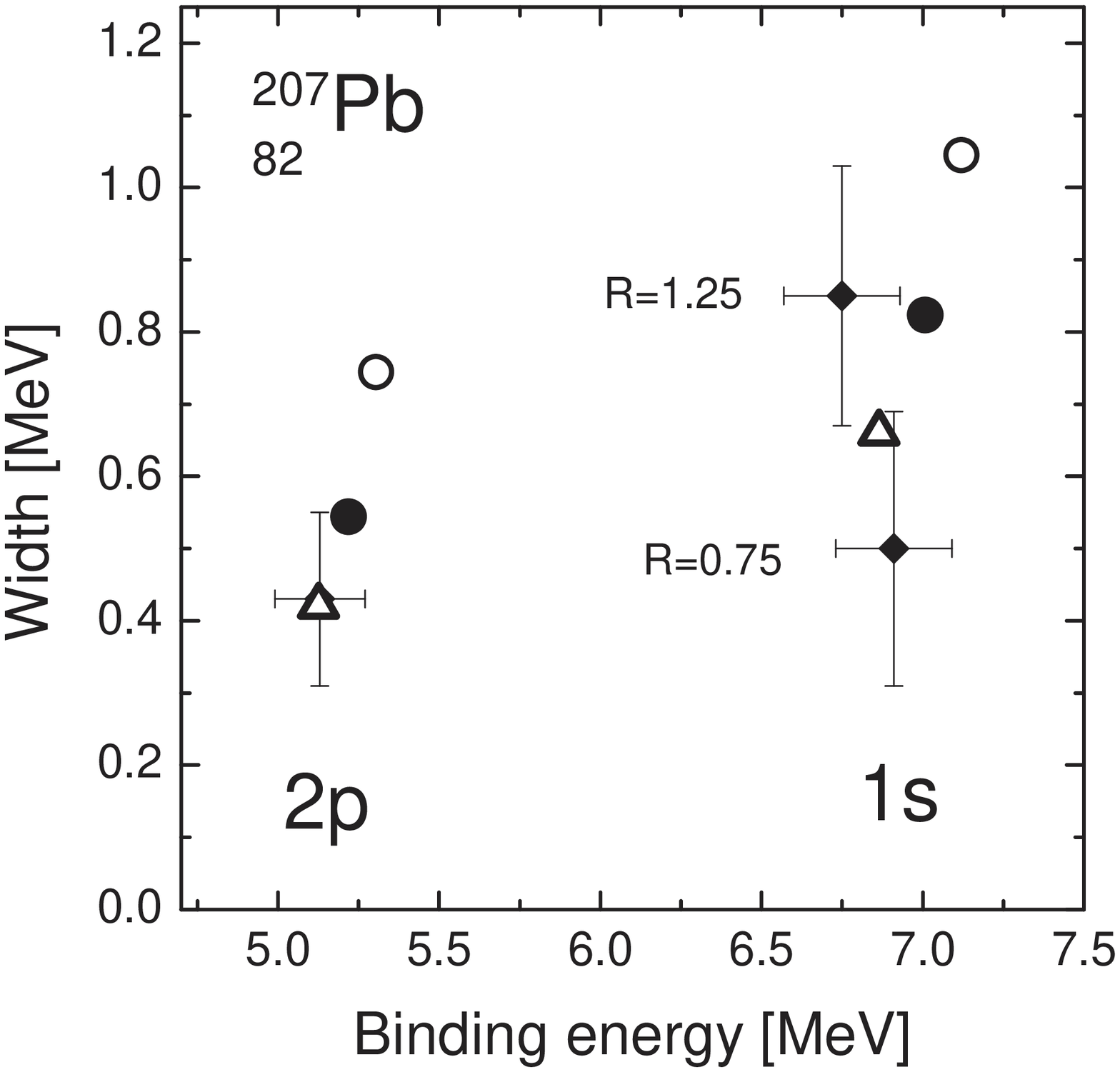}
\quad
\includegraphics[width=5.5cm,clip=true]{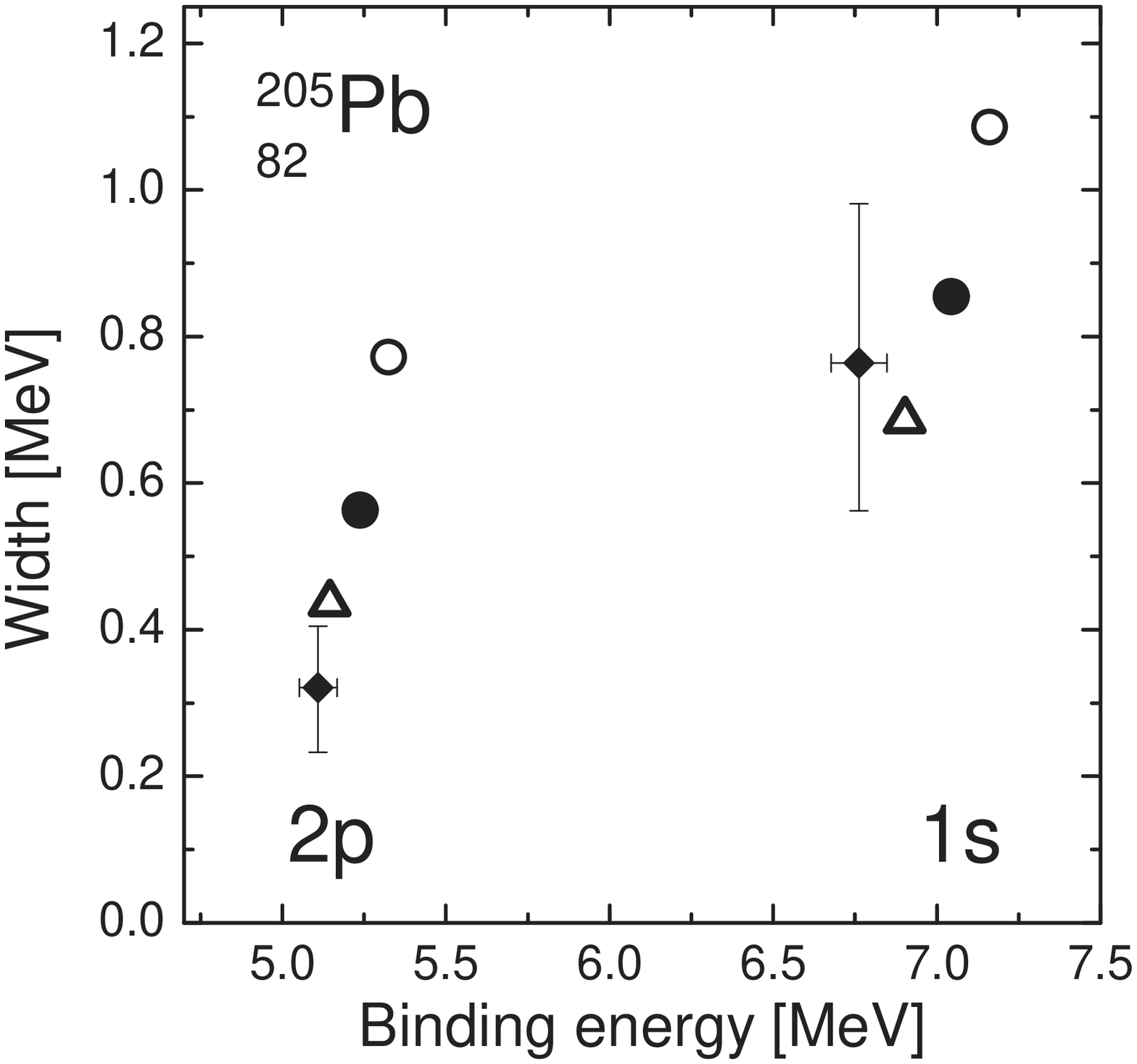}
\end{center}
\caption{Binding energies and widths of deeply bound pionic states
in the isotopes $^{207}$Pb (left figure) and $^{205}$Pb (right
figure). Diamonds show the experimental data from~\cite{pb207}.
Uncertainties in the extraction of the $1s$ level for $^{207}$Pb
are indicated by different choices of a control parameter $R$ as
specified in~\cite{pb207}. The results for the polarization
operator (\ref{pistat},\ref{pidyn}) are depicted by open and
filled circles, respectively. Triangles show the results obtained
with the chiral polarization operator (\ref{pichi}).}
\label{fig:pb}
\end{figure*}

Solutions of the wave equation (\ref{KGE}) for Pb isotopes
with the energy independent (threshold) input (\ref{pistat})
for the pion-nuclear optical potential are shown in
Fig.~\ref{fig:pb} by open circles. The filled circles in
Fig.~\ref{fig:pb} are the results obtained with the polarization
operator
\be
\label{pidyn}
\Pi(\om)=\Pi_{\rm phen}(\om; \rho_p(r),\rho_n(r))\,,
\ee
in which we keep the explicit energy
dependence as given by the driving terms (\ref{tpl}).
The energy dependence effects are evidently important, moving the
calculated results closer to the data. Indeed, with the gauge invariant
introduction of the electromagnetic interaction in the pion
polarization operator (via the replacement $\om\to \om -V_c(r)$)
the off-shell pion-nucleon  scattering amplitudes are probed at
energies $\om-V_c(r)> m_\pi$. This increases the repulsion in
$T^-(\omega)$ and disbalances the cancelation between the
sigma-term $\sigma_N$ and  the range term $-\beta\omega^2$ in
$T^+(\omega)$, giving $T^+(\om -V_c(r)) <0 $. Omitting the
replacement $\om\to \om-V_c(r)$ in $\Pi(\om)$, we would have
$\om<m_\pi$, and this would reduce the repulsion in $T^-(\omega)$
and turn on attraction in $T^+(\omega)$,  thus
leading in the wrong direction. Taking both the energy dependence
and the proper gauge invariant substitution via $\Pi(\om-V_c(r))$
is therefore an essential ingredient.

After these qualitative considerations we proceed now to the systematic
calculation of the pion polarization operator using
in-medium chiral perturbation theory. Here we extend the results
of ref.~\cite{Kaiser01} at the two-loop level by taking into
account the explicit (off-shell) energy dependence.
The polarization operator has the form:
\be
\label{pichi} \Pi_{\rm }(\om)= \Pi_0(\om)+
\Pi_{\rm ds}(\om)+\Pi_{\rm rel}(\om)+\Pi_{\rm cor}(\om)\,.
\ee
The first term corresponds to the linear density approximation:
\be
\Pi_0(\om)=
\parbox{15mm}{\includegraphics[width=15mm,clip=true]{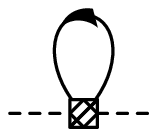}}
=-T^-(\om)\,\delta \rho-T^+(\om)\,\rho\,.
\ee
The isospin-even off-shell $\pi N$-scattering amplitude
at zero pion momentum can be written in the following
form (for $\omega > m_\pi$):
\be \nonumber
T^+(\om)=\frac{\sigma_N-\beta\omega^2}{f_\pi^2}+ \frac{3g_A^2
m_\pi^3}{16\pi f_\pi^4}
+ {3
g_A^2 Q^2 m_\pi \zeta \over 64 \pi f_\pi^4}+ i T_{\rm im}
\ee
where $\beta=g_A^2/4M-2c_2-2c_3$\,,
$\sigma_N = -4c_1 m_\pi^2-9g_A^2 m_\pi^3/64\pi f_\pi^2$
and $T_{\rm im}= {\omega^2 Q}/({8\pi f_\pi^4})$\,. The
nucleon axial-vector coupling constant has the value $g_A=1.27$.
We introduce the
abbreviation $Q=\sqrt{\omega^2-m_\pi^2}$.
The second-order low-energy constants $c_{1,2,3}$ (for notations
see ref.~\cite{BKM}) are tuned to the empirical values of the
sigma-term~\cite{sainio}, $\sigma_N=45$~MeV, and the $\pi N$ scattering
length, $T^+(m_\pi)=0$\,.

The parameter $\zeta$ reflects freedom in the
choice of the interpolating pion field in the effective chiral
Lagrangian~\cite{Wirzba02,Park01}. It enters all interaction
vertices with three and more pions. The one-loop correction to the
(off-shell) pion self-energy in vacuum depends also on this
parameter $\zeta$. By requiring that the residue at the pion pole
remains equal to one~\footnote{We discard here the fact the wave
function renormalization factor of the pion in vacuum receives
also the contribution $-2l_4 m_\pi^2/f_\pi^2$ from the low-energy
constant $l_4$.} as it is implicit in the form of the KG
equation~(\ref{KGE}), one gets the constraint $\zeta=0$.

The isospin-odd off-shell $\pi N$-amplitude at zero momentum reads:
\be
\nonumber
T^-(\om)&=& \frac{\om}{2 f_\pi^2} \bigg[1+{\gamma
\,\om^2 \over (2\pi f_\pi)^2} \bigg]
\\ \label{Tmn}
&-&\frac{\omega^2 Q}{8\pi^2f_\pi^4}
\ln\frac{\omega+ Q}{m_\pi}+\frac{i}{2}\,T_{\rm im}
\ee
with $\gamma= (g_A\pi f_\pi/M)^2
+\ln(2\Lambda/m_\pi)$\,. The cut-off scale $\Lambda=737\,$MeV $\simeq 8 f_\pi$ is
chosen to reproduce the central empirical value of the on-shell
scattering amplitude at threshold $T^-(m_\pi) = 1.85\pm0.09\,$~fm~\cite{Schroeder99}.
We  neglect here small additional counter-term contributions proportional to
the third order low-energy constants $\bar d_j$ of ref.~\cite{nadja}.

The next term in (\ref{pichi}) corresponds to the important
Ericson-Ericson double scattering correction~\cite{ee66}
generalized to isospin asymmetric nuclear matter and off-shell
pions:
\be\nonumber
&&\Pi_{\rm ds}(\om)=
\parbox{18mm}{\includegraphics[width=18mm,clip=true]{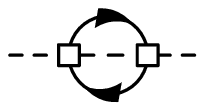}}
=\frac{\om^2}{3(4\pi f_\pi)^4} \Big\{ L(k_p,k_p,Q)
\\ \label{ds}
&&\qquad\qquad +L(k_n,k_n,Q)+2L(k_p,k_n,Q)\Big\}\,,
\ee
where
$k_{p,n}= (3\pi^2 \rho_{p,n})^{1/3}$ refer to the proton and
neutron Fermi momenta and
\be
\nonumber L(k_p,k_n,Q)&=&4k_p k_n(Q^2 + 3k_p^2 +3 k_n^2 )  \\
\nonumber &&+ 8Q(k_n^3 -k_p^3) \ln{ Q+k_n-k_p \over Q-k_n+k_p}  \\
\nonumber && - 8Q(k_p^3 +k_n^3)\ln{ Q+k_p+k_n \over Q-k_p-k_n} \\
\nonumber && +\Big[ 3( k_p^2 -k_n^2)^2 + 6Q^2(k_p^2 + k_n^2 )-
Q^4\Big]
\\
&& \times\ln{(k_n- k_p)^2-Q^2\over ( k_p +k_n)^2-Q^2} \,. \ee

The third term in (\ref{pichi}) is a small relativistic
correction from the particle-hole (Born) diagram evaluated at zero pion
momentum:
\be\label{pirel}
\Pi_{\rm
rel}(\om)=\parbox{20mm}{\includegraphics[width=20mm,clip=true]{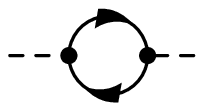}}
=\frac{g_A^2\om}{10(\pi M f_\pi)^2}(k_p^5-k_n^5)\,.
\ee
The last term in (11) represents the effect induced by
$\pi\pi$-interactions with two virtual pions being absorbed on the
nucleons in the Fermi-sea, and by an additional two-loop
correction~\cite{Park01,Kaiser01}:
\be\nonumber
&&\Pi_{\rm cor}(\omega)=
\parbox{30mm}{\includegraphics[width=30mm,clip=true]{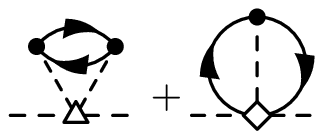}}
= \frac{g_A^2}{20(4\pi f_\pi)^4}
\\ \nonumber
&&\times \left\{(Q^2(\zeta+2)+5m_\pi^2)
[H(k_p,k_p)+ H(k_n,k_n)]
\right.\\ \label{picor} \label{cor} && \left.
\quad+ Q^2(8\zeta-4) H(k_p,k_n) \right\}\,.
\ee
The function $H(k_p,k_n)$ consists of the last four terms
written in eq.~(12) of ref.~\cite{Kaiser01}. In the actual calculation the
contribution $\Pi_{\rm cor}(\omega)$ turns out to be negligibly
small.

Solutions of the wave equation (\ref{KGE}) for $1s$ and $2p$
levels using  (\ref{pichi}) are shown
in Fig.~\ref{fig:pb} by triangles.
They agree well with experimental data.
We also find that the energy dependent polarization operator
(\ref{piphen},\ref{teff}) gives  equally good results as
(\ref{pichi}) if the amplitude $T^-(\om)$ in (\ref{tpl}) is extended
to include the $\om^3$ term in (\ref{Tmn}), with the
parameter $\gamma$ tuned to the empirical value of the scattering
length.

\begin{figure}
\begin{center}
\includegraphics[width=5.5cm,clip=true]{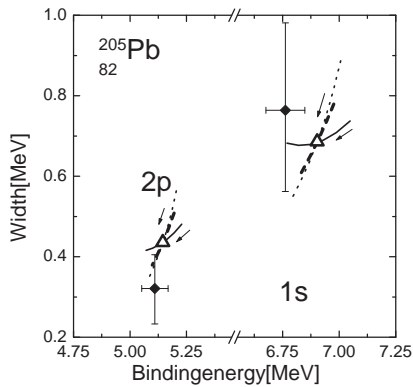}
\end{center}
\caption{ Dependence of energies and widths of  pionic levels for
$^{205}$Pb on parameters of the optical potential.  Solid lines
correspond to the variation of the neutron radius in the interval
$\delta R_n=-0.2 \div 0.2$~fm. Dashed lines  show the variation
with the pion nucleon sigma term $\sigma_N(0)=20\div 65$~MeV.
Dotted lines represent the variation of $\Re B_0=(-0.5\div 0.5)\,
\Im B_0$. Arrows indicate the directions in which  results move as
the varied parameters increase. Triangles and experimental points
are the same as in Fig.~\protect\ref{fig:pb}.  } \label{fig:pbvar}
\end{figure}

In Fig.~\ref{fig:pbvar} we examine the dependence of our
results for $^{205}$Pb on the less constrained parameters of the
model.  Variations of the neutron radius $R_n$
affect mainly the binding energy, whereas the sigma term
$\sigma_N$ and  $\Re B_0$ have a stronger impact on the level
width. Note the strong correlation in the effects induced by
changes of $\Re B_0$ and $\sigma_N$\,.

In summary, we have demonstrated that the long standing issue of
the "missing repulsion" in the s-wave pion nucleus potential can
be naturally resolved by taking into account the explicit energy
dependence of the pion self-energy and the gauge invariant
incorporation of electromagnetic interactions. The experimental
data for $1s$ and $2p$ levels in Pb isotopes are well reproduced.
We have also clarified that, to leading order, the energy
dependence effects can be interpreted in terms of the in-medium
reduction  of the pion decay constant, or equivalently, the
renormalization of the intrinsic pion wave function in matter.

We thank E. Friedman, P. Kienle, M. Lutz, E. Oset, D. Voskresensky
and T. Yamazaki for discussions and communications. We also
gratefully acknowledge valuable contributions by R.~Leisibach
at the early stages of this work. This work is supported in part by BMBF and
GSI.

\end{document}